\documentstyle[preprint,aps,epsfig]{revtex}
\draft

\begin{document}

\title{Microstructure of the highly dense MgB$_2$ superconductor by 
       transmission electron microscope}

\author{Gun Yong Sung\footnote{Correspondence should be addressed to Gun Yong Sung (e-mail: gysung@etri.re.kr )}, Sang Hyeob Kim, and JunHo Kim}
\address{Telecommunications Basic Research Laboratory, 
Electronics and Telecommunications Research Institute, Taejon 305-350, Rep. of Korea}

\author{Dong Chul Yoo, Ju Wook Lee, and Jeong Yong Lee}
\address{Department of Material Science and Engineering,
Korea Advanced Institute of Science and Technology, Taejon 305-701, Rep. of Korea}

\author{C. U. Jung, Min-Seok Park, W. N. Kang, Du Zhonglian, and Sung-Ik Lee}
\address{National Creative Research Initiative Center for Superconductivity
and Department of Physics, Pohang University of Science and Technology,
Pohang 790-784, Rep. of Korea}

\maketitle

\begin{abstract}

The microstructure of the MgB$_2$ superconductor sintered at high temperature under a high 
pressure of $\sim$ 3 GPa was investigated by using a high-resolution transmission electron 
microscope (HRTEM).  The TEM images did not show any pores in the specimen.  All grains 
were compactly connected, and no discernable empty spaces or impurities at the boundaries 
existed over the regions investigated. The HRTEM image showed clearly each constituent atom 
that formed the basal hexagonal plane, without any defect in a single grain. The a-axis lattice 
parameter, 0.307 nm, from this direct measurement was shorter than the value, 0.314 nm 
obtained from samples prepared using diffusion techniques.  A minor impurity phase, which 
was most probably MgB$_4$ and did not form interfacial layers was also observed, but was well 
isolated from the main MgB$_2$ phase.  Our results verify that the MgB$_2$ powder was sintered 
under high temperature and high pressure into its theoretical density without any porosity or 
grain growth. 
\end{abstract}


Superconductivity at about 40 K in MgB$_2$ was announced quite recently.\cite{1} The $T_c$ of this 
binary intermetallic superconductor is the highest among the non-oxide and non-C$_{60}$-based 
compounds and is near the upper limit suggested by the conventional BCS theory.\cite{2} A phonon-
mediated paring mechanism has been proposed to explain the superconductivity and is 
supported by the observation of a boron isotope effect with an isotope exponent 
$\alpha_B \sim 0.26$.\cite{3,4}
The carrier type was predicted to be positive,\cite{5} which was confirmed by Hall measurements.\cite{6} 

Typically, MgB$_2$ powders have been synthesized by sintering mixtures of Mg and B powders 
in an inert atmosphere.\cite{3,7,8} Some have reported the synthesis of high-density pellets sintered 
under high pressures of several GPa; this high density makes it easier to study the transport 
properties.\cite{6,9,10} A strong connection between the grains, which increases with increasing 
sintering temperature under high pressure, has been suggested to explain the concurrent 
changes in the resistivity and the low-field magnetization.\cite{11} The microstructure of high-density 
MgB$_2$ has been observed by using a scanning electron microscope (SEM)\cite{10,11} or an optical 
microscope.\cite{7}  However, the high-resolution transmission electron microscope (HRTEM) images 
of MgB$_2$ showing the detailed structure of a single grain and the grain boundary have not yet 
been reported. 

In this paper, we report HRTEM observations of the microstructure of the high-density 
MgB$_2$ superconductor.  The TEM images showed that no pores existed in the specimen.  All 
grains were well connected without any impurities at the grain boundaries, much like at the 
twin boundary of YBa$_2$Cu$_3$O$_y$.  The HRTEM image showed each constituent atom clearly, and 
not a single MgB$_2$ grain containing any impurities. An a-axis lattice parameter of 0.307 nm was 
obtained from this direct measurement and is shorter than the value of a = 0.314 nm obtained 
from samples prepared using a diffusion techniques.  The minor impurity phases were well 
isolated from the main MgB$_2$ phase.  Our results suggest that the MgB$_2$ powder was sintered 
into its theoretical density without any impurities at the grain boundaries. These properties can 
make the transport measurements very reliable. 

A 12-mm cubic multi-anvil-type press was used for the high-pressure sintering.\cite{9} The 
starting material was a commercially available powder of MgB$_2$. The pressed pellet was placed 
in a Au capsule in a high-pressure cell and pressurized up to 3 GPa.  While the pressure was 
maintained, the heating temperature was increased linearly; then, it was kept at 900$^\circ$C for 1 
hour. The samples were then quenched to room temperature. The pellets, weighing about 130 
mg, were about 4.5 mm in diameter and 3.3 mm in height.  A TEM specimen was prepared by 
mechanical polishing with a Tripod polisher for precise thickness control. A specimen was 
polished on diamond abrasive films down to about a 10 $\mu$m thickness and then ion-milled with 
Ar ions for 20 hours.  This was done using a Gatan dual ion miller at 5 kV and 0.6 mA.  A 
high-resolution image, a bright field image, and a selected area electron diffraction (SAED) 
pattern were obtained with a transmission electron microscope (JEOL JEM-2000EX) operating at 
200 kV.

\begin{figure}
\center
\epsfig{figure=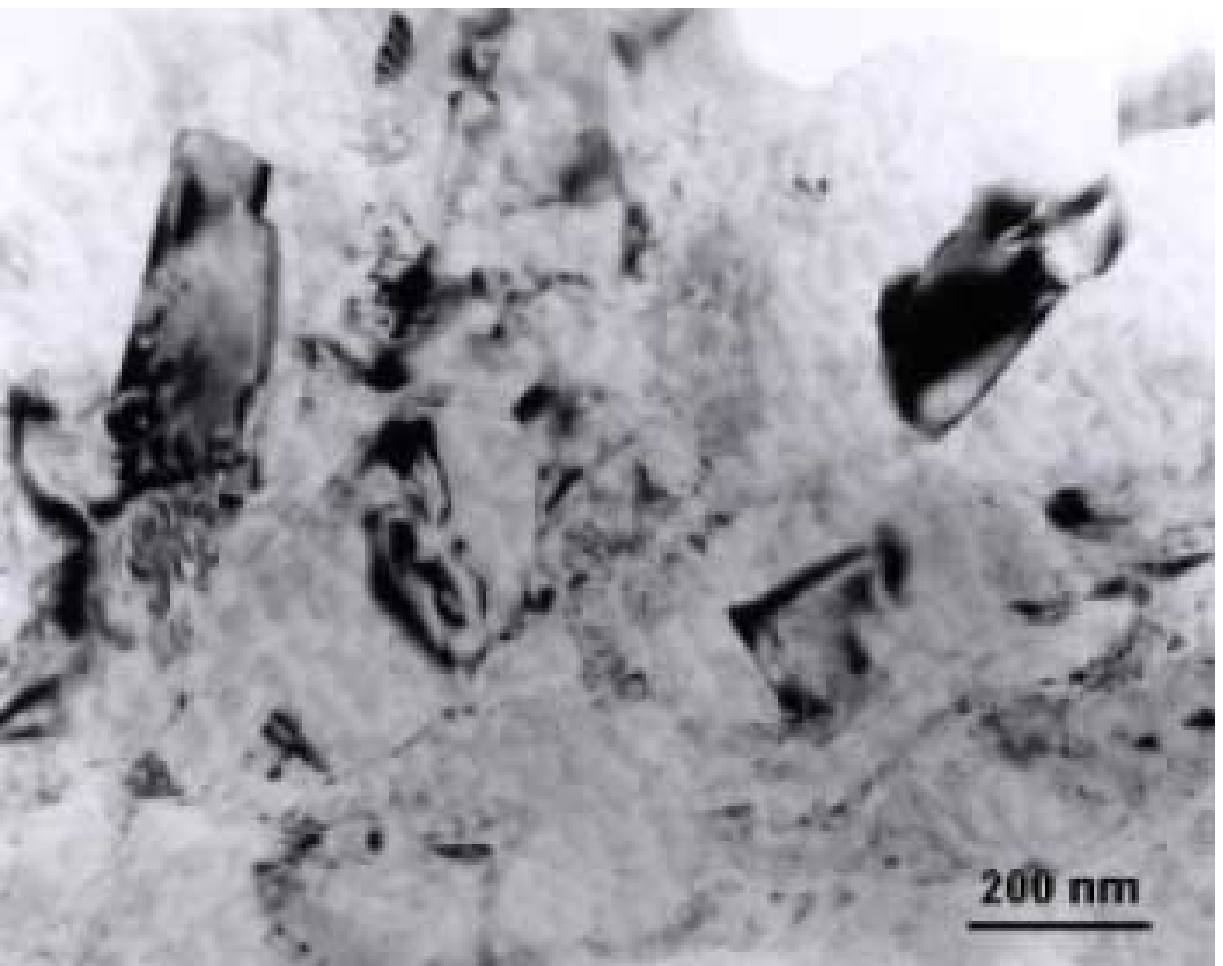,width=0.65\textwidth,height=0.30\textheight,clip=,angle=0}
\caption{TEM image of the highly dense MgB$_2$ observed at low magnification}
\label{fig1}
\end{figure}

Figure \ref{fig1} shows the microstructure observed using the TEM at low magnification. 
Surprisingly, the microstructure showed that the sample was pore-free and was almost 100\% 
densified. The grain sizes were in the range of 300 nm to 800 nm. All grains were well connected 
without any impurities at the grain boundaries. Only the crystal orientation was different from 
grain to grain at the boundary, much like the case of the twin boundary of YBa$_2$Cu$_3$O$_y$.  The 
dark and the light regions in Fig. \ref{fig1} are from grains with different crystallographic orientations. 
Some elongated grains with aspect ratios of 4 to 6 were also observed.

\begin{figure}
\center
\epsfig{figure=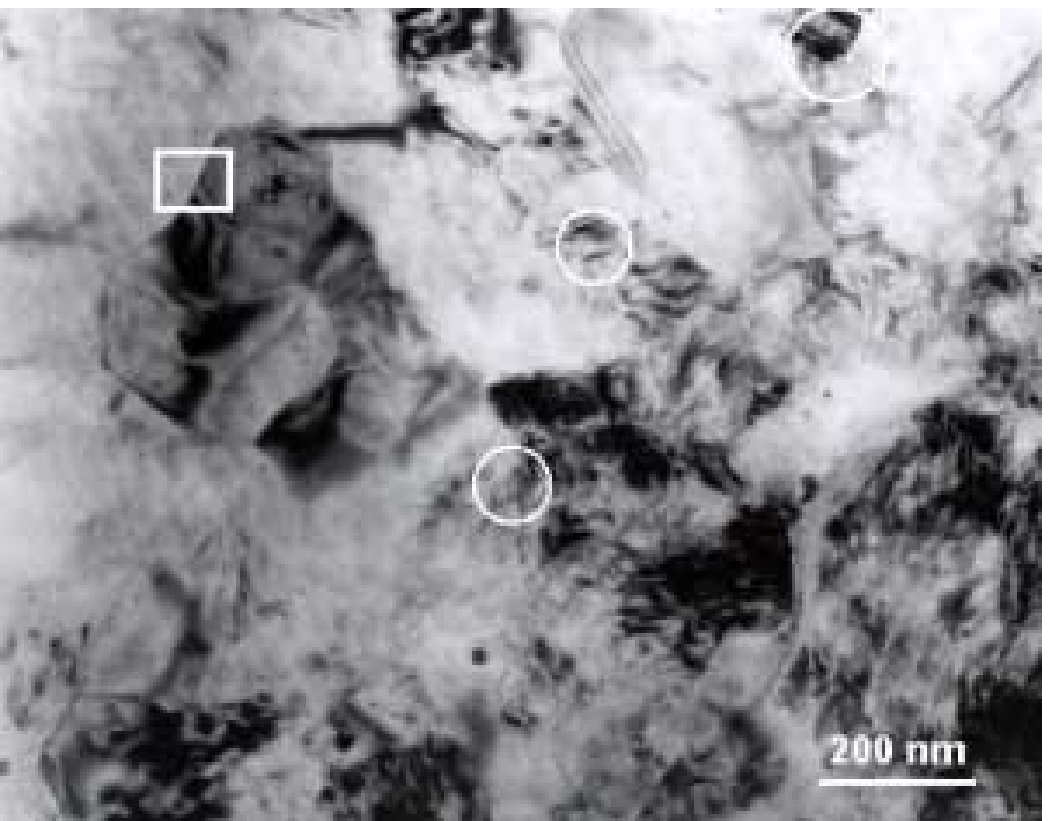,width=0.65\textwidth,height=0.30\textheight,clip=,angle=0}
\caption{TEM image of the highly dense MgB$_2$ observed at low magnification}
\label{fig2}
\end{figure}

Figure \ref{fig2} shows a bright-field image, mostly consisting of MgB$_2$. Over small areas with sizes 
less than 100 nm, lattice fringes with an interplanar spacing larger than that of MgB$_2$ were 
observed in the regions denoted by circles. Previously, two kinds of impurities, cubic MgO and 
orthorhombic MgB$_4$, had been reported by Takano {\it et al}. for MgB$_2$, sintered at high pressure.\cite{10}  In 
that case, a trace of MgO was present in the starting MgB$_2$ powder, and the small amount of 
MgB$_4$ originated from the partial decomposition of MgB$_2$ to MgB$_4$ at 1000 ¨¬C under a pressure of 
3.5 GPa.\cite{10}  Since the c-axis lattice constants of MgO and MgB$_4$ are 0.4213 nm (JCPDS 04-0829) 
and 0.7472 nm (JCPDS 15-0299), respectively, the observed impurities seemed to be mostly MgB$_4$ 
because of the large c-axis constant.

\begin{figure}[ccc]
\center
\epsfig{figure=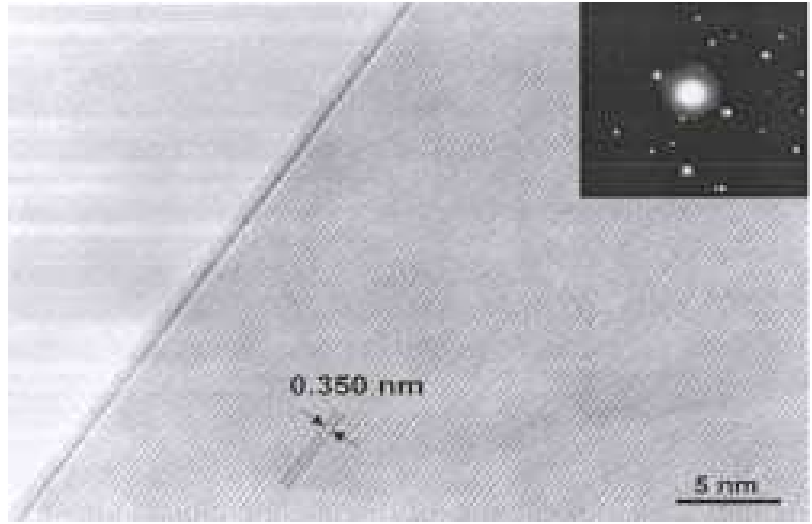,width=0.65\textwidth,height=0.30\textheight,clip=,angle=0}
\caption{HRTEM image of the area marked in the square box shown in Fig. 2}
\label{fig3}
\end{figure}

Figure \ref{fig3} shows the HRTEM image of the grain boundary marked by the white square in Fig. 
2. There is no visible intergranular phase at this grain boundary.  The beam direction of the 
SAED pattern in the inset of Fig. \ref{fig3} was [010], and the defect-free lattice fringe, which is due to 
the (001) basal hexagonal plane of MgB$_2$. And its interplanar spacing, equivalent to the c-axis 
lattice parameter, was estimated to be 0.350 nm, which is same as the reported value.\cite{3}

\begin{figure}[ccc]
\center
\epsfig{figure=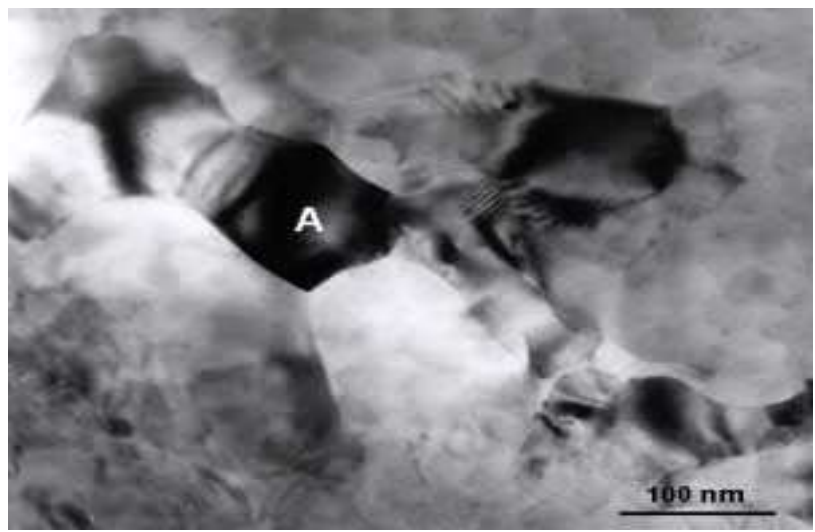,width=0.65\textwidth,height=0.30\textheight,clip=,angle=0}
\caption{TEM image of the highly dense MgB$_2$ observed at high magnification}
\label{fig4}
\end{figure}

\begin{figure}[ccc]
\center
\epsfig{figure=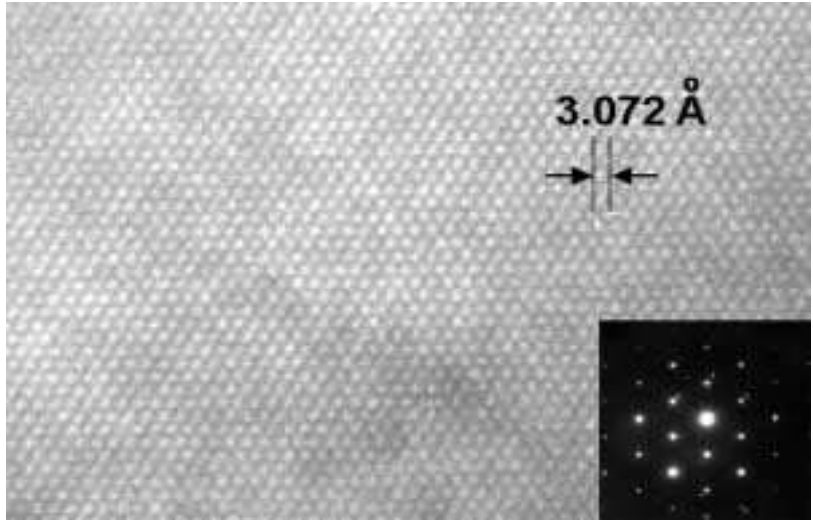,width=0.65\textwidth,height=0.30\textheight,clip=,angle=0}
\caption{HRTEM image of the MgB$_2$ grain marked as A in Fig. 4}
\label{fig5}
\end{figure}

Figure \ref{fig4} shows a bright-field image at high magnification. Neighter porosity Nor 
intergranular phase, were observed at the grain boundaries, as in Fig. \ref{fig1}.  The MgB$_2$ grain 
marked A was selected for HRTEM imaging, and that image is shown in Fig. \ref{fig5}.  The beam 
direction of the SAED pattern from the grain, shown in the inset of Fig. \ref{fig5}, was [001], and the 
HRTEM image of the grain shows a very clean basal plane consisting of an Mg atomic array.  
The in-plane a-axis lattice parameter was estimated to be 0.307 nm, which is shorter than the 
value of a = 0.314 nm for samples prepared using a diffusion techniques.

In summary, we used a HRTEM to obtain the microstructure of the high-density MgB$_2$ 
superconductor.  No porosity and no intergranular phases were observed on the TEM images 
of the polycrystalline MgB$_2$.  The observed minor impurity phases, which seemed to be MgB$_4$, 
were well separated from the main MgB$_2$ phase. The lattice fringe image from the HRTEM 
image shows well-aligned atomic array, without any defects in a single grain, and proves that 
this sample is very clean. Thus, the measurements of the transport properties for our MgB$_2$ 
sintered under high pressure should not be affected by impurities or defect scattering.\cite{12}

This work was supported by the Korea Research Council for Industrial Science \& Technology and the Ministry of Science and Technology through the Creative Research Initiative Program and the National Research Laboratory Program.

\end{document}